# Comets in Australian Aboriginal Astronomy

Duane W. Hamacher[1] and Ray P. Norris[1]

[1]Department of Indigenous Studies, Macquarie University, NSW, 2109, Australia
Corresponding author e-mail: duane.hamacher@mq.edu.au

**Abstract**

We present 25 accounts of comets from 40 Australian Aboriginal communities, citing both supernatural perceptions of comets and historical accounts of bright comets. Historical and ethnographic descriptions include the Great Comets of 1843, 1861, 1901, 1910, and 1927. We describe the perceptions of comets in Aboriginal societies and show that they are typically associated with fear, death, omens, malevolent spirits, and evil magic, consistent with many cultures around the world. We also provide a list of words for comets in 16 different Aboriginal languages.

**Keywords**: Comets, Aboriginal Australians, Ethnoastronomy, History of Astronomy.

## 1.0    Introduction

Cometography is the study of comets from both from a scientific and historic perspective (Kronk, 2003a). Recorded sightings of comets date back to the second century BCE, with the possibly earliest written recording of a comet by the Chinese in 1059 BCE (Yeomans & Kiang, 1981; Xu et al, 2000:107-125). While historic accounts of comets and their role in mythology have been widely described in the literature (e.g. Baillie & McCafferty, 2005; Burnham, 2000; Donnelly, 2005; Kronk, 2003a,b; Levy, 1994; Schechner, 1999), little research has been conducted regarding Aboriginal Australian accounts of comets. Ethnographic literature on Aboriginal communities reveals several perceptions of comets and accounts of historic comets. In some cases the comet is described but the author does not specify to which comet the account is referring (e.g. Morrill, 1864:61; Roth, 1984:8), although the description and dates provided allow us to identify the most probable comet to which the account is referring. In other cases, perceptions of a comet are described, but do not correspond to any particular one. The description of the object, combined with the dates of the vent, allow us to identify the particular comet discussed, providing a more complete historical account. Evidence shows that Aboriginal Australians were astute astronomers (see Norris & Hamacher, 2009), having a complex social and religious economy associated with the night sky. Similar studies relating to meteors and cosmic impacts in Aboriginal Astronomy are presented in Hamacher & Norris (2010; 2009), respectively.

This paper presents various Aboriginal perceptions of comets and well as descriptions of historic bright comets, including C/1843 D1, C/1844 Y2, C/1861 N1, C/1901 G1, 1P/1909 R1, C/1910 A1, and C/1927 X1. We begin by introducing the reader to basic information about Aboriginal Australians, the concept of the Dreaming, and a description of comets as a celestial phenomenon. We then present our data collection methods and describe the various perceptions of comets, then dividing up identifiable comets by the date of the recorded account.

### 1.1    Aboriginal Cultures

The cosmic and terrestrial landscapes were an inseparable and integral component of daily life to the hundreds of Aboriginal communities in Australia and played a vital role in the structure and evolution of oral traditions and ceremonies. Celestial objects and phenomena were believed to be intricately tied to events on the earth. Because of this, rare cosmic events, such as a bright meteor, comet, or eclipse, had great significance and meaning to those people who witnessed them. These events were often recorded in oral tradition and art, and were integral to the preservation and dissemination of both general and sacred knowledge within the community.

When researchers and scholars began visiting and studying Aboriginal communities and recording oral traditions in the early 19[th] century, very little was known about the diverse Aboriginal hunter/gatherer societies that not only survived, but thrived in some of the world's harshest climates for over 40,000 years. After the British colonised Australia in the late 18[th] century, Aboriginal communities were devastated by disease, limited access to resources, and outright genocide. As a result, the total Aboriginal population reduced from over 300,000 in 1788 (Jupp, 2001:93) to just 93,000 by 1900 (Australian Bureau of Statistics, 2002). In many places, especially near British colonies, entire cultures, including the language, customs, and oral traditions, were virtually destroyed.

Despite the damage to many Aboriginal cultures, several Aboriginal communities still thrive and retain their collective knowledge and traditions, while a substantial amount of information about other Aboriginal communities has been recorded in the literature. Although this information represents a tiny fraction of the knowledge originally possessed by Aboriginal communities, it can help non-Aboriginal people understand Aboriginal beliefs and perceptions of natural phenomena at the time they were recorded.

**1.2    The Dreaming**

Common among most Aboriginal cultures is the concept of 'The Dreaming', an English term coined by Francis Gillen in 1896 and adopted by Spencer & Gillen (1899) to refer to the period in the religious oral traditions of the Northern Arrernte people of Central Australia (Dean, 1996). The Dreaming is not a universal, homogeneous concept spanning all Aboriginal groups, but instead possesses substantial variations in the way it is viewed and understood by various Aboriginal groups across Australia. The Dreaming is viewed by some Aboriginal groups (e.g. the Tiwi and Wiradjuri) as the period during the creation of the world when totemic ancestors came into being, representing a past reality. For other groups, it represents a past, current and future reality, either concurrently parallel to our own reality (e.g. the Ooldea and Warrabri), or within our own reality (e.g. the Murinbata and Mardudjara). It should also be emphasized that the Dreaming does not necessarily represent a linear progression of time. In some cases, such as during ritual ceremonies, the past can become the present, so the term "Dreamtime" used in an all-encompassing sense is not accurate, as it denotes a linear timeline, separating past, present, and future. The Dreaming is part of a diverse and complex social structure and system of laws and traditions that has been an integral aspect of Aboriginal cultures for thousands of years (see Bates, 1996; Rose, 1996, 2000; Stanner, 1965, 1976, 1979).

**2.0    Methodology**

The available literature, including books, journals, magazines, audio and video sources were reviewed for references to comets[1] or descriptions that may refer to comets but do not explicitly identify them as such. Most of the data are taken from ethnographies collected in the 19[th] and early 20[th] centuries. 41 accounts were collected, including 16 Aboriginal words for comets, representing 40 Aboriginal groups from all Australian states except Tasmania (eight form Victoria, Queensland and the Northern Territory, seven from Western Australia, six from South Australia, and four from New South Wales). Historically visible comets are identified by the date and or description of the account. In some cases, the identification is clear, while in other cases, the identification is inferred and should not be considered definitive.

We must stress that Indigenous views and accounts of comets are not homogeneous or unchanging over time. It is common for a particular celestial object or phenomena to have different views among members of the community, and these views are dynamic and evolving. It is therefore essential to understand that the accounts provided in this paper simply reference the community from which they came, and does not imply that everyone within that community or language group had the same view, reaction, or association with comets. Statements such as "The [Aboriginal group] saw comets as ______" are used only to denote the Aboriginal group or geographic area from which the account was taken. We must also emphasise that these records come largely from Western researchers and scholars, including some colonists that were not trained in linguistics, ethnology or anthropology. Therefore, these accounts are inherently biased. Fredrick (2008) analyses the academic backgrounds of the major contributors to Australian Aboriginal Astronomy, which gives insight into the reliability and accuracy of the information provided by these researchers.

**3.0    Results**

**3.1    Aboriginal Perceptions of Comets**

When Westerners began studying Aboriginal communities in the early 1800s, they noted that the Aboriginal people viewed extraordinary or unusual natural events with great dread (e.g. Eyre, 1845:358-359; Palmer, 1884:294). The unexpected arrival of a bright comet often triggered fear and were associated with death, spirits, or omens – a view held by various cultures around the world (e.g. Andrews, 2004; Bobrowsky & Rickman, 2007; McCafferty & Baillie, 2005). Such views include those of the Tanganekald of South Australia who perceived comets as omens of sickness and death (Tindale, n.d.), the Mycoolon of Queensland who greeted comets with fear (Palmer, 1884:294), the Kaurna of Adelaide who believed that the sun father, called *Teendo Yerle*, had a pair of evil celestial sisters who were "long" and probably represented comets (Clarke, 1997:129; Schurmann, 1839) and the Euahlayi of New South Wales saw comets as evil spirits that drank the rain-clouds causing drought[2], with the cometary tail representing a large thirsty family that drew the river into the clouds (Parker, 1905:99). The Moporr clan of Victoria described a comet as *Puurt Kuurnuuk* - a great spirit (Dawson, 1881:101), while the Gundidjmara of Victoria saw a comet as an omen that lots of people will die (Howitt and Stähle, 1881). Aboriginal people in the Talbot District of Victoria likened comets (called "*Koonk cutrine too*") to smoke, where "*too*" means "*to smoke*" (Smyth, 1878:200). This is similar to a report from Cape York Peninsula, where an Aboriginal community saw a comet as the smoke of a campfire (Roth, 1984:8). Similarly, the Aboriginal people of Bentinck Island in the Gulf of Carpentaria called a comet *burwaduwuru*, which means "testicle with smoke" (Evans, 1992:196). A common view among Aboriginal communities of the Central Desert links comets to spears (e.g. Spencer, 1928:409). A Pitjantjatjara man named Peter Kunari (Anon, 1986:20) described comets as the manifestation of a being named *Wurluru* who lived in the sky and carried spears that he occasionally threw across the heavens (a possible reference to meteors?). A similar association is shared by the Kaitish, which discussed further in Section 3.3. The Rainbow Serpent, a much-feared evil spirit found in the Dreamings of many Aboriginal groups, was sometimes associated with comets (e.g. Healy, 1978:194)[3]. Trezise (1993:107) speculated that the origins of the Rainbow Serpent lay in transits of Halley's comet, which was seen every 76 years, reinforcing stories handed down by Kuku-Yalanji law-carriers and custodians of the Bloomfield River, Queensland.

Spencer & Gillen (1899:550; 1904:627-628; 1927:415–417) describe a form of evil magic called Arungquilta, which involved meteors and produced comets and was used to punish unfaithful wives in Arunta communities. If a woman ran away from her husband, he would summon men from his group and a medicine man to perform a ceremony intended to punish her. In the ceremony, a pictogram of the woman was drawn in the dirt in a secluded area while the men chanted a particular song. A piece of bark, representing the woman's spirit, was impaled with a series of small spears endowed with Arungquilta and flung into the direction of where they believed the woman to be, which would appear in the sky as a comet (bundle of spears). The Arungquilta would find the woman and deprive her of her fat. After the emaciated woman died, her spirit appeared in the sky as a meteor. Strehlow (1907:30) cites a nearly identical ceremony. However, in Strehlow's account, the man felt pity for his wife and decided to revive her by rubbing fat into her body. As she healed, the comet faded from view. In some Arrernte and Luritja communities, comets are spears thrown by an ancestral hero to make his wife obedient to him (Strehlow, 1907:30). To some Arrernte clans, a comet was also a sign that a person in a neighbouring community had died, usually because of infidelity, and pointed to the direction of the deceased (Spencer & Gillen, 1899:549). A similar description is given by Piddington (1932:394) about the Karadjeri of coastal Western Australia, but is instead attributed to meteors. Given the two accounts by Spencer and Gillen of the same ceremony, it is possible they are confusing comets with meteors.

A direct association between comets and death is highlighted by a story from the Kimberleys of a great flood that was brought on by a "star with trails" called *Kallowa Anggnal Kude* (Mowaljarlai & Malnic, 1993:194). Bryant (2001; et al, 2007:210-211) contends this account is a description of a comet impact in the Indian Ocean off the northwest coast of Western Australia, which he speculates caused a massive tsunami that devastated the region. Bryant speculates that the "star with trails" is depicted in a rock painting at a place called 'Comet Rock' near Kalumburu, Western Australia (home the Wunambal and Kwini), which lies on a plain 5 km from the sea that is covered in a layer of beach sand.

### 3.2     Historic Visible Comets

In much of the literature, ethnographic accounts do not simply describe perceptions of comets, but often refer to particular historic comets, including the "Great Comets" (bright comets) of 1843, 1861, 1901, 1910, 1927, and Comet Halley.  In some cases, the comet is not identified by name, but is inferred from the description and date.  Details of each comet observation or description are presented below.

### 3.2.2     C/1843 D1 (Great March Comet of 1843)

The Great Comet of 1843 (C/1843 D1) was a bright, sun-grazing comet visible in the Southern skies from late-February to mid-April.  It was visible near the sun (within one degree) and became brightest on 07 March (see Kronk, 2003a:129; Sekanina & Chodas, 2008).  The comet, so frightening in its brilliance (see Figure 1), prompted Aboriginal people near Port Lincoln, South Australia to run and hide in caves (Schurmann, 1846:242).  The Ngarrindjeri of South Australia saw the comet as a harbinger of calamity, specifically to the white colonists.  They believed the comet would destroy Adelaide then travel up the Murray causing havoc in its path, as described by Eyre (1845:358-359)[4]:

> *In March 1843, I had a little boy living with me* [in Moorunde, SA] *by his father's permission, whilst the old man went up the river with the other natives to hunt and fish. On the evening of the 2nd of March a large comet was visible to the westward, and became brighter and more distinct every succeeding night.  On the 5th I had a visit from the father of the little boy who was living with me, to demand his son; he had come down the river post haste for that purpose, as soon as he saw the comet, which he assured me was the harbinger of all kinds of calamities, and more especially to the white people. It was to overthrow Adelaide, destroy all Europeans and their houses, and then taking a course up the Murray, and past the Rufus* [the site of an Aboriginal massacre]*, do irreparable damage to whatever or whoever came in its way. It was sent, he said, by the northern natives, who were powerful sorcerers, and to revenge the confinement of one of the principal men of their tribe, who was then in Adelaide gaol, charged with assaulting a shepherd; and he urged me by all means to hurry off to town as quickly as I could, to procure the man's release, so that if possible the evil might be averted. No explanation gave him the least satisfaction, he was in such a state of apprehension and excitement, and he finally marched off with the little boy, saying, that although by no means safe even with him, yet he would be in less danger than if left with me.*

Le Souëf (in Smyth, 1878:296) recounted events that took place when the Great Comet of 1843 was seen in Victoria (specific location not identified).  When it was first seen, it caused "dreadful commotion and consternation" among the communities.  "Spokesmen [presumably Elders or medicine men] gesticulated and speechified far into the night" in an attempt to rid the comet, but with no success.  When their actions seemed in vain, they packed-up camp in the middle of the night and moved to the other side of the river and remained huddled together until morning.  They believed that the comet had been sent to them by the Aboriginal people near Ovens River in northern Victoria to cause harm.  They left the area and did not return until the comet faded away.  Aboriginal people near Kilmore, Victoria told Curr (1886:50) that the tail of this "grand comet" consisted of spears thrown by Aboriginal people near Goulburn to one another[5].

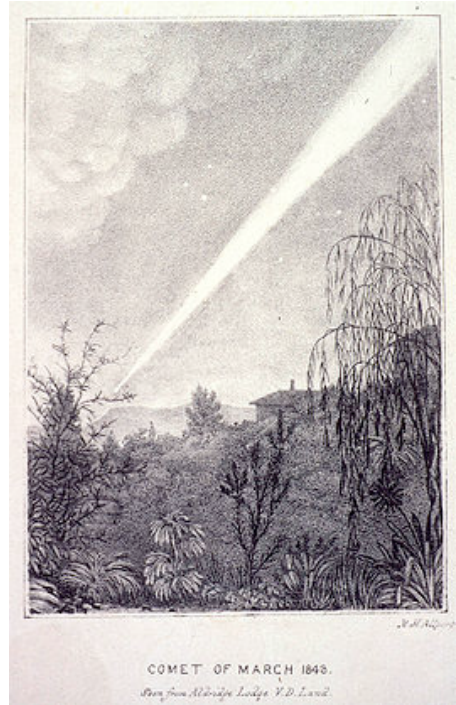

**Figure 1:** The Great Comet of 1843 (C/1843 D1) as seen from Tasmania (Van Diemen's Land). Painting by Mary Morton Allport (1806-1895). Reproduced from Wikimedia Commons under Creative Commons License.

### 3.2.3    C/1844 Y2  (Comet Wilmot)

There were no Aboriginal accounts of this comet in the reviewed literature. However, the explorer Ludwig Leichhardt saw Comet Wilmot in the sky on 29 December 1844 while walking along the banks of a creek in central Queensland (Lang, 1847:315), prompting him to name the site Comet Creek. The town of Comet, Queensland (originally Cometville) takes its name from this creek (now called Comet River, Edwards, 1994).

### 3.2.4    C/1861 N1 (Comet Tebbutt)

After surviving the shipwreck of the *Peruvian* in the Great Barrier Reef in 1846, four members of the crew reached Cleveland Bay on the coast of Queensland near present-day Townsville. One of the survivors, James Morrill, lived among the local Aboriginal people for 17 years, publishing his journals in 1864 (Morrill, 1864). He notes how the Aboriginal people used the same word for comets and stars (*nilgoolerburda*) and explains that comets were believed to be the spirits of men killed far away returning home, making their way from the clouds to the horizon. He described seeing a comet during the previous dry season (June to November) and noted that the Aboriginal people thought it was "one of the tribe who had been killed in war" (*ibid*:61). Morrill does not give a date of his sighting, but does go on to say that he witnessed a nearly total solar eclipse about six years earlier. From 1846 to 1864, only two nearly total solar eclipses (where the moon covered over 80% of the sun) were visible from this region: 05 April 1856 ($t_e$ = 17:05:31) and 18 September 1857 ($t_e$ = 17:28:04, where $t_e$ is the time of mid-eclipse), despite Morrill claiming that he only saw one eclipse during his time living among the Aboriginal people. This gives a period of approximately five years between the eclipse and the comet sighting, revealing the best candidate is Comet Tebbutt (C/1861 N1, see Figure 2). Comet Tebbutt was discovered in Australia and visible from mid-May through mid-August, during the dry season (see Kronk, 2003a:293; Orchiston, 1998a; 1998b), which implies that Morrill's account was recorded in 1862. Comet Tebbutt would have appeared brightly in the northern sky throughout June with the tail extending below the horizon (see Ellery, 1861; Raynard, 1872; Scott, 1861), which may explain why the Aboriginal people told Morrill the comet was a spirit coming down from the clouds onto the horizon (Morrill, 1864:61).

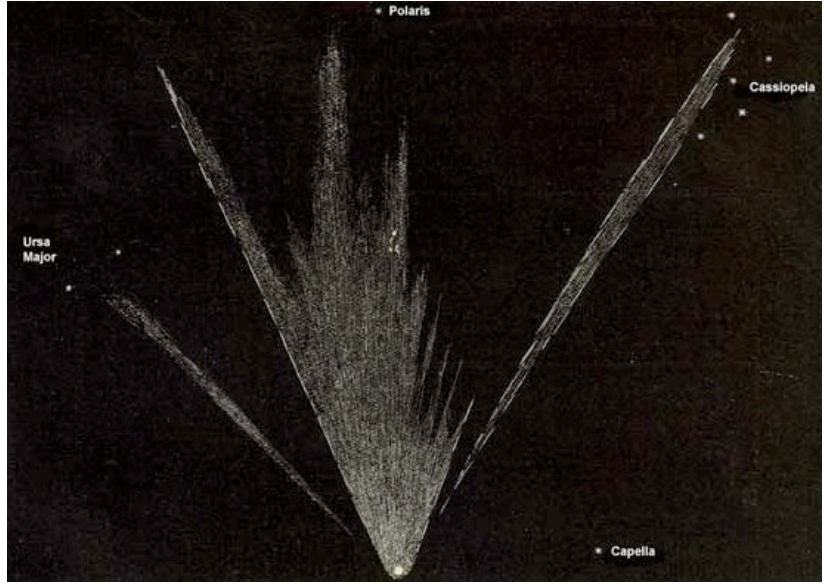

**Figure 2:** A drawing of Comet Tebbutt (C/1861 N1) made on 30 June 1861, drawn with respect to a Northern Hemisphere observer. The image would appear upside-down to observers in the Southern Hemisphere. Reproduced from Wikimedia Commons under Creative Commons License.

### 3.2.5    C/1901 G1 (Comet Viscara)

While engaging in ethnographic fieldwork in Queensland from 1901 to 1908, anthropologist and Northern Protector of Aboriginals, Walter E. Roth (1984:8), noted that the Tjungundji people near Mapoon (Marpuna) on the western coast of Cape York Peninsula saw a comet as a fire lit by two old women. This was probably a reference to the recent Great Comet of 1901, which was visible exclusively in the southern skies from mid-April to late-May and displayed distinctive, bright twin tails (Gill, 1901; Tebbutt, 1901; see Figure 3). In early May, during its brightest peak, the comet transversed the boundary between Taurus and Eridanus. The head of the comet, of magnitude 0 on 03 May and +2 on 06 May, would have appeared in the western evening skies near the horizon with the twin tails, comprising a 30 degree straight tail and a 10 degree curved tail, pointing upwards towards the star Sirius. By 12 May 1901, the longer tail extended to the star δ Leporis (Kronk 2003b; Observatory, 1901). This would have looked very much like two smoke columns diverging from a single point on the horizon. The comet remained visible until 23 May with the tails increasing in length to 45 and 15 degrees, respectively (Bortle, 1998).

The Kaitish of the Northern Territory believed a comet was a bundle of spears belonging to a star endowed with a very strong magic. The people feared these spears would be thrown to earth, killing many. Spencer & Gillen (1904:629; 1912:327) describe a bright comet visible during their stay in 1901, which is probably a reference to C/1901 G1. To avert the evil of the comet, a young, celebrated medicine man named Ilpailurkna was visiting the area from the neighboring Unmatjera clan. Each night he would project his magic stones towards the comet. As the comet faded away, its evil was overcome and the people were very grateful that Ilpailurkna had saved them[6]. In the eyes of the community, had Ilpailurkna not driven the comet away, it would have fallen to earth as a bundle of spears and everyone would have been killed (*ibid*:630).

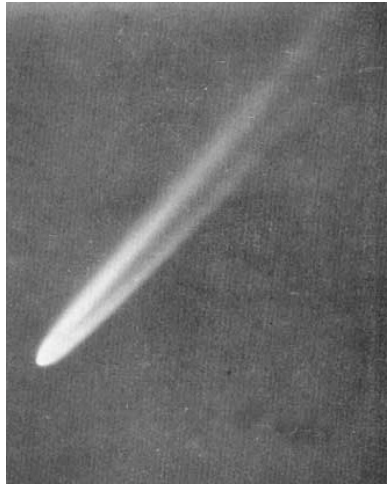

**Figure 3:** The Great Comet of 1901 with bright twin tails, taken from the Royal Observatory, Cape of Good Hope, South Africa (Mitton, 2009:120).

### 3.2.6    C/1910 A1 (Great Daylight Comet of 1910) and 1P/1909 R1 (Comet Halley)

In 1910, the world awaited the return of the famous Comet Halley in May.  However, the unexpected arrival of a bright comet in mid-January created much fear and awe (e.g. New York Times, 1910; Burnham, 2000:184).  Deemed the Great Daylight Comet of 1910 (see Figure 4), it was bright enough to be seen during the day and at its peak, was brighter than Venus.  It began to fade away in early February, followed a few months later by the arrival of the fainter, but still significant, Comet Halley (Kronk 2003b).  When Comet Halley returned in 1986, many of the older people around the world who recalled seeing it in 1910 had clearly described the Great Daylight Comet of 1910 and not Halley[7] (*ibid*).

In 1985 Jack Butler, a Jiwarli man from the Henry River in Western Australia, told of a "star with a tail in the east" he saw early in the year 1910 as a child (Butler & Austin, 1986:85-88).  The comet caused fear among the elder men who "questioned what it was".  When the comet faded away, then men were confused and wondered where it had gone.  According to Butler, the object he saw in 1910 was Comet Halley.  However, the Great Daylight Comet of 1910 was prominent in the morning twilight, consistent with the "star with a tail in the east" visible early in the year.  Therefore, it is probable that Butler was describing the Great Daylight Comet of 1910 rather than Comet Halley[8].

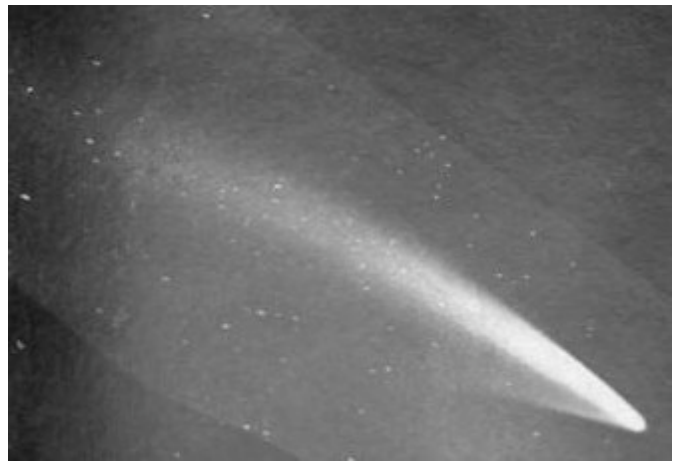

**Figure 4:** The Great Daylight Comet of 1910 appeared just four months before Comet Halley but was brighter than Venus at its peak (image rotated clockwise by 90°). Photograph from Lowell Observatory,

### 3.2.7   C/1927 X1 (Comet Skjellerup-Maristany)

Paddy Roe, a Nyigina elder, told of the appearance of a comet in the early 20th century by an Aboriginal community on the Roebuck Plains west of Broome, Western Australia (Duwell & Dixon, 1994:80). The comet, which he described as a "star with a tail", was seen as a bad omen. However, after nothing bad happened, the community held a celebratory corroboree. Roe states that the comet was first seen during the "new moon when the moon was a crescent" (this refers to the time after a new moon when the moon appears as a thin crescent). These accounts date to the period "between the Wars", presumably referring to World Wars I and II (between 1918 and 1939). The best candidate is the Great Comet of 1927 (see Figure 5), co-discovered on 04 December 1927 by the Australian amateur astronomer John Francis Skjellerup when it was a third magnitude object with a one-degree tail (although others had claimed to have discovered it on 28 November 1927; Orchiston, 1999). The comet, first visible in the Southern Hemisphere, was visible primarily during the day and early evening. By the time it was visible at night, it faded rapidly. Since the comet was near the solar disc but still visible during the day, the sighting of Comet Skjellerup-Maristany is consistent with being seen at the time of a new moon (the day it was claimed to have first been discovered, 28 November 1927, was just after new moon, see Makemson, 1928; Seargent, 2009:147-148), although this identification is uncertain.

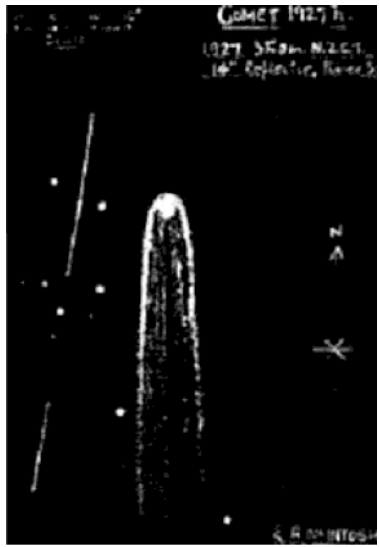

**Figure 5:** Drawing of Comet Skjellerup-Maristany by R.A. McIntosh, 5 December 1927. Image reproduced courtesy of Wayne Orchiston.

### 3.3   Aboriginal Names for Comets

In Table 1, we present a list of Aboriginal names for comets, citing 16 Aboriginal groups from all states except Tasmania.

**Table 1**: List of Aboriginal names for comets and the sources from which they were taken.

| Group | State | Term | Source |
|---|---|---|---|
| Bindal | QLD | Nilgoolerburda | Morrill (1864:61,62) |
| Boiwoorarng | VIC | Jajowerong | Smyth (1878:177) |
| Gumbaybggirr | NSW | Gumugan | Morelli (2008:160) |
| Gunditjmarra (Moporr) | VIC | Puurt Kuurnuuk | Dawson (1881:101) |
| Kayardild | QLD | Burwaduwuru | Evans (1992:196) |
| Kwini | WA | Kallowa Anggnal Kude | Mowaljarlai & Malnic (1993:194) |
| Ngiyampaa | NSW | Yangki (Comet Halley) | Thieberger & McGregor (1983:2.8) |
| Parnkalla | SA | Yandarri | Schurmann (1844:79) |
| Pitjantjatjara | NT | Wuuluru | Goddard (1992:202) |
| Djadjawurung[1] | VIC | Koonk cutrine too | Smyth (1878:200) |
| Wiradjuri | NSW | Muma | Rudder (2005:403) |
| Wunambal | WA | Kallowa Anggnal Kude | Mowaljarlai & Malnic (1993:194) |
| Yarra Yarra | VIC | Bullarto tutbyrum | Smyth (1878:136) |
| Yolngu | WA | Ngarrpiya | Lowe (2004:116) |
| Unspecified | WA | Binnar (also meteor) | Moore (1842:126) |
| Unspecified | VIC | Boiwoorarng | Smyth (1878:162) |

[1] From the Talbot District, Victoria (language group name not specified in text). Language group taken from the AIATSIS map of Aboriginal Languages.

## 4.0 Discussion and Conclusion

The relatively sudden and effectively unpredictable nature of comets (that is, unpredictable without making detailed observations over long periods of time) are the likely driving force behind their generally negative views not only among Aboriginal Australians, but among most cultures of the world (see Ridpath, 1985). Of the 25 accounts given in this paper, all but two were attributed to negative concepts, namely fear, bad omens, death, malevolent spirits, or evil magic. This is consistent with global views of comets (e.g. Ridpath, 1985; Yeomans, 1991). The only non-negative views of comets likened them to smoke (Bobrowsky & Rickman, 2007; Evans, 1992:196; Roth, 1984:8; Smyth, 1878:200), a view shared by some Maori tribes of New Zealand, who called some comets *Auahi-roa* or *Auahi-turoa*, from the words "*auahi*" meaning "smoke" and "*roa*" meaning "long" (see Best, 1922; Orchiston, 2000) and the Aztecs of Mesoamerica, who called comets "*citlalinpopoca*", meaning "star that smokes" (Aveni, 1980:27).

It is unclear whether comets had always been viewed with fear or whether this fear was triggered by a coincidental catastrophic event. Clearly, some accounts establish a perceived link between the appearance of a comet and unrelated malign events, such as drought (e.g. Parker, 1905:99), death or disease (e.g. Howitt & Stähle, 1881; Spencer & Gillen, 1899:549; Tindale, n.d.), the presence of a hostile enemy (e.g. Morrill, 1864:61), or a natural disaster (e.g. Mowaljarlai & Malnic, 1993:194) – views shared by many cultures of the world (e.g. Andrews, 2004:111-121; Köhler, 1989:292)[9]. While scientists now know that comets are responsible for a percentage of destructive exploding meteors[10] (see Napier & Asher, 2009) and cosmic impacts (see Bobrowsky & Rickman, 2007), there is no evidence to link comets with disease outbreaks.

Most recorded views of comets indicate that the people who saw them were surprised. Although comets are not seen as frequently as other transient celestial phenomenon (such as meteors), they do make an appearance every few years. Eclipses occur less frequently than comets, but appear as a re-occurring phenomenon in the oral traditions of many Aboriginal communities (e.g. Bates, 1944; Johnson, 1998; Norris & Hamacher, 2009; Warner, 1937). However, there are few accounts of comets in oral traditions (at least to the extent that they can be easily identified as such) and we are curious as to why this is the case.

Are there accounts of comets that we have failed to recognise? Some of these accounts may be found in the form of rock art, such as motifs found in rock engravings of the Sydney region, including the Bulgandry figure near Woy Woy, NSW (see Figure 6). If the objects held by Bulgandry represent the sun and crescent moon (e.g. Norris, 2008), then we may speculate that his "hair" actually represents a comet. The hair or headdress of some culture heroes, such as Daramulan (McCarthy, 1989), has a similar appearance to comets and have been described by Elkin (1949:131) as representing spears, suggesting a possible parallel with the communities from the Northern Territory that associate comets with spears. Numerous other examples of similar motifs are found in rock art of the Sydney region. We are currently looking into the possibility that these engravings represent comets, but any connection at present is simply speculation.

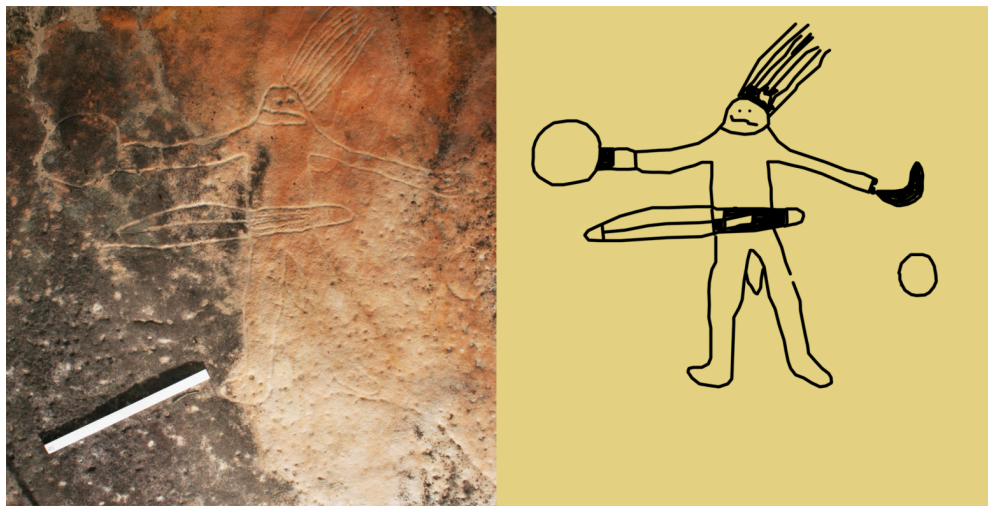

**Figure 6:** The Bulgandry petroglyph near Woy-Woy, NSW.  **Left** - Image © Ray Norris (2007).  **Right** - Drawing by W.D. Campbell (1893).

Although nearly all of the descriptions in this paper are second-hand Western accounts of Aboriginal perceptions of comets, these accounts provide an important historical record of Great Comets from an Aboriginal perspective.  We conclude that comets are frequently associated with spears due to the comet's appearance resembling a bundle of spears, and that perceptions of comets amongst Aboriginal societies were usually associated with fear, death, omens, malevolent spirits, and evil magic, due to their awe-inspiring and relatively unexpected nature, consistent with many cultures around the world.  We attribute their generally negative views to their unpredictable and significant appearance in an otherwise well-ordered cosmos.  However, we remain puzzled by the fact that nearly all accounts are from colonial times, with few accounts in the recorded oral tradition and no apparent trace of comets in pre-colonial Aboriginal art.

### Acknowledgements

The authors would like to thank and acknowledge the Aboriginal elders and custodians of Australia, the Darug People (the Traditional Custodians of the land on which Macquarie University is situated), Wayne Orchiston, The Royal Observatory (Cape of Good Hope, South Africa), John Clegg, and Kristina Everett.  Hamacher was funded by the Macquarie University Research Excellence Scholarship (MQRES) within the Department of Indigenous Studies at Macquarie University.

### Notes

[1] Comets and meteors are sometimes conflated.  Tindale (1983:376) categorizes comets and fireballs (bright meteors) together, despite them being different phenomena, though he only discusses the former.  Some languages use the same word for both phenomena.  For example, although the Spanish words for comet and meteor are *cometa* and *meteoro*, respectively, the word *cometa* is preferred by rural Mexicans to describe both phenomena (Köhler, 1989:289).  In some Australian Aboriginal languages, the word for comet and meteor are reported to be the same, such as *nilgoolerburda,* in the Bindel language of northern Queensland (Morrill, 1864:61) and *binnar* in a Western Australian language (Moore, 1842:126,145).  In some cases, the description of one seems to indicate the other.

[2] It is unclear if this view was due to the coincidental arrival of a comet before or during a major drought.  One candidate is the Great Comet of 1825 (C/1825 N1), which was visible from late August until the end of December 1825.  From 1826-1829, a severe drought hit New South Wales, causing Lake George and the Darling River to completely dry up (Shaw, 1984).  Another possible candidate (of many) was the Great

Southern Comet of 1880, visible in the evening skies in February (Kirkwood, 1880; Morris, 1880), which preceded a significant drought in New South Wales.

[3] Additional information regarding the evil of comets can be found in Barker (n.d.) at the Australian Institute for Aboriginal and Torres Strait Islander Studies in Canberra. This item is under restricted access and cannot be copied or quoted (Fredrick, 2008:105).

[4] Johnson (1998:49) mistakenly attributes this account to the Great Comet of 1811 (personal communication).

[5] Curr stated that he recalled the event in 1842, although the only bright comet over the years prior to and following 1842 was in 1843.

[6] A similar description of medicine men throwing magical stones at a comet to drive it away is given by Hambly (1936:23).

[7] There are no reported Aboriginal accounts of the 1986 return of Comet Halley found in the literature. However, the comet was adopted as the logo for the Arnhem Land Progress Aboriginal Corporation (2009). Comet Halley's return is also featured in Aboriginal artwork and literature, including Brogus Nelson Tjakamurra's painting 'Halley's Comet' (1986) and Sam Watson's novel 'The Kadaitcha Sung' (1990).

[8] During the joint University of California-Los Angeles/University of Adelaide Expedition into northwestern Australia (1953-1955), Tindale (1983:377) explained how researchers used Comet Halley as an indicator of age for more mature Aboriginal informants. While using either Comet Halley or the Great Comet of 1910 would have been sufficient for their study, it is probable that the informant's descriptions were of the January comet.

[9] Throughout history, people have tried to make a connection between passing comets and destructive events (e.g. Gadbury, 1665), including disease outbreaks (e.g. Bobrowsky & Rickman, 2007: Chap 5) and natural disasters (Bryant, 2001). Comet impacts may have caused environmental change in the past, creating poor environmental conditions where starvation and the spread of disease were more rampant. Others (e.g. Wickramasinghe et al, 2004) have speculated that cometary debris contains microbial bacteria, referred to as Cometary Panspermia, which seeded life on earth and may be responsible for disease epidemics, such as SARS and the Bubonic Plague. This idea has met substantial criticism (see Vaidya, 2009) and is not generally accepted by the scientific community. Scientists, however, more generally accept the hypothesis that amino acids and water were brought to earth via comets, which later evolved into life.

[10] Although the composition of the Tunguska (1908, see Napier and Asher, 2009), Curuçá (1930, see Bailey et al, 1995), and Guyana (1935, see Steel, 1996) bolides have not been well established, they all occurred when the earth passed through major meteoroid streams, which are produced by the dust tails of passing comets.

Duane Hamacher is a PhD candidate in the Department of Indigenous Studies at Macquarie University. After graduating in physics from the University of Missouri and obtaining a master's degree in astrophysics from the University of New South Wales, he was awarded a Research Excellence Scholarship to study Aboriginal Astronomy at Macquarie. Duane is also an astronomy educator at Sydney Observatory and the Macquarie University Observatory and Planetarium.

Professor Ray Norris is an astrophysicist at CSIRO Astronomy & Space Science and an Adjunct Professor in the Department of Indigenous Studies at Macquarie University. While obtaining his MA and PhD in physics from Cambridge and Manchester, respectively, he began researching the archaeoastronomy of British stone arrangements. Ray is the Secretary of the International Society of Archaeoastronomy and Astronomy in Culture (ISAAC) and heads the Aboriginal Astronomy Research Group, working with Indigenous groups such as the Wardaman and Yolngu communities of the Northern Territory.